\documentclass[showpacs,amsmath,twocolumn,showkeys,prl]{revtex4-1}
\usepackage{dcolumn}    
\usepackage{bm}         
\usepackage{ifpdf}
\usepackage{amssymb,lineno,amsfonts}
\usepackage{graphicx}   
\usepackage{dcolumn}    
\usepackage{bm}         
\usepackage{bbm}
\usepackage{mathrsfs}
\usepackage{upgreek}
\usepackage{mathtools}
\usepackage{epstopdf}
\usepackage{setspace}
\usepackage{hyperref}
\usepackage[matrix,frame,arrow]{xypic}
\usepackage{rotating}
\usepackage{float}
\usepackage{graphicx}
\usepackage{natbib}
\usepackage{xcolor}
\definecolor{med-blue}{RGB}{25,25,112}
\hypersetup{colorlinks, linkcolor={red},citecolor={blue}, urlcolor={blue}}

\input{Qcircuit}
\begin{document}

\title{Violation of Entropic Leggett-Garg Inequality in Nuclear Spins}
\author{Hemant Katiyar$^1$, Abhishek Shukla$^1$, Rama Koteswara Rao$^2$, and T. S. Mahesh$^1$}
\email{mahesh.ts@iiserpune.ac.in}
\affiliation{$ ^1 $Department of Physics and NMR Research Center,\\
Indian Institute of Science Education and Research, Pune 411008, India \\
$ ^2 $Department of Physics and NMR Research Center,
Indian Institute of Science, Bangalore, India}

\begin{abstract}
We report an experimental study of recently formulated entropic Leggett-Garg inequality (ELGI)
by Usha Devi \textit{et al.} (arXiv: 1208.4491v2 (2012)). 
This inequality places a bound on the statistical measurement outcomes of dynamical
observables describing a macrorealistic system.  Such a bound is not necessarily
obeyed by quantum systems, and therefore
provides an important way to distinguish quantumness from classical behavior.   
Here we study ELGI using a two-qubit nuclear magnetic resonance system. 
To perform the noninvasive measurements required for the ELGI study, we prepare the system qubit in a
maximally mixed state as well as use the `ideal negative result measurement' procedure with 
the help of an ancilla qubit.  The experimental results show a clear violation of ELGI
by over four standard deviations.  These results agree with the predictions of quantum theory.
The violation of ELGI is attributed to the fact that certain joint
probabilities are not legitimate in the quantum scenario, in the sense they do not
reproduce all the marginal probabilities.  Using a three-qubit system, we experimentally
demonstrate that three-time joint probabilities do not reproduce certain two-time marginal
probabilities.
\end{abstract}

\maketitle

\textit{Introduction.---}
The behavior of quantum systems is often incomprehensible by classical notions, the
best examples being nonlocality \cite{epr,bell} and contextuality \cite{contextual}.
Quantum systems are \textit{nonlocal} since they violate Bell's inequality, which 
assumes that local operations on one of the two space-like separated objects can not disturb
the measurement outcomes of the other \cite{bellineq}.  
The quantum systems are also \textit{contextual} in the sense that a measurement outcome
depends not only on the system and the property being measured, but also on
the context of the measurement, i.e., on the set of other compatible properties which are
being measured along with.

Another notion imposed on classical objects is macrorealism, which is 
based on two criteria: 
(i) the object remains in one or the other of many possible states at all times, and
(ii) the measurements are noninvasive, i.e., they reveal the state of the object without disturbing 
the object or its future dynamics.
Quantum systems are incompatible with these criteria and therefore violate
bounds on correlations derived from them.
For instance, Leggett-Garg inequality (LGI) sets up macrorealistic bounds 
on linear combinations of two-time correlations of a dichotomic observable
belonging to a single dynamical system \cite{lgi}.
In this sense, LGI is considered as a temporal analogue of Bell's inequality.
Quantum systems do not comply with LGI,
and therefore provide an important way
to distinguish the quantum behavior from macrorealism.
Violations of LGI by quantum systems have been investigated and 
demonstrated experimentally in various systems
\cite{opto,*deco,*e-trans,*back,*quantumdot,*crystal, Palacios,*photons,*semiweak, maheshnmr,*souza,noninvasive}.

For understanding the quantum behavior it is important to investigate
it through different approaches, particularly
from an information theoretical point of view.
For example, an entropic formulation for Bell's inequality has 
been given by Braunstein and Caves \cite{bc},
and more recently that for contextuality has been given independently by Rafael and Fritz 
\cite{fritz} and Kurzy\'{n}ski {\it et.al.} \cite{kurz}.
Recently, an entropic formulation of LGI has also been introduced  
by Usha Devi \textit{et al.} \cite{usha}, 
in terms of classical Shannon entropies associated 
with classical correlations.  Such entropies obey certain
constraints, which when violated would imply non-existence of 
legitimate joint probabilities (JP) for all the measurement outcomes. 

Here we report an experimental demonstration of violation of entropic LGI (ELGI) 
in an ensemble of spin $1/2$ nuclei using nuclear magnetic resonance (NMR) techniques.
Although NMR experiments are carried out at a high temperature limit, the nuclear spins
have long coherence times, and their unitary evolutions can be 
controlled in a precise way.
The large parallel computations carried out in an NMR spin ensemble
assists in efficiently extracting the single-event probability (SEP) and JP.
The simplest ELGI study involves three sets of two-time joint measurements of 
a dynamic observable belonging to a `system' qubit at time instants
$(t_1,t_2)$, $(t_2,t_3)$, and $(t_1,t_3)$.  The first measurement in each
case must be `noninvasive' in the sense, it should not influence the outcome
of the second measurement.  These noninvasive measurements (NIM) can be 
performed with the help of an ancilla qubit.

Further, it has been argued in \cite{usha} that
the violation of ELGI arises essentially due to the fact 
that the JP do not originate from a legitimate grand probability (of 
which the JP are the marginals).
Here we describe extracting three-time JP using a three-qubit system, and
demonstrate experimentally that it can not reproduce all
the marginal probabilities (MP) substantiating this feature.

In the following we briefly revisit the theory of ELGI \cite{usha}
and then we describe the circuits for the measurement 
of SEP and JP. Later we detail its experimental study using a two-qubit NMR system.  Then we describe
the study of three-time JP using a three-qubit NMR system.

\textit{Theory.---}
Consider a dynamical observable $Q(t_k)=Q_k$ measured at different time instances $t_k$.
Let the measurement outcomes be $ q_k $ with probabilities $ P(q_k) $.
In classical information theory, the amount of information stored 
in the random variable $Q_k$ is given by the Shannon entropy \cite{chuangbook},
\begin{eqnarray}
H(Q_k)=-\sum_{q_k} P(q_k)\log_2{P(q_k)}.
\label{shne} 
\end{eqnarray}
The conditional information stored in $ Q_{k+l} $ at time $ t_{k+l} $,
assuming that the observable $Q_k$ has an outcome $ q_k $,
is 
\begin{eqnarray}
H(Q_{k+l}\vert Q_k=q_k) = 
-\sum_{q_{k+l}} P(q_{k+l}\vert q_k)\log_2P(q_{k+l}\vert q_k) \nonumber,
\end{eqnarray}
where $P(q_{k+l}\vert q_k)$ is the conditional probability.
Then the mean conditional entropy is given by,
\begin{eqnarray}
H(Q_{k+l}\vert Q_k) &=& -\sum_{q_k} P(q_k)H(Q_{k+l}\vert Q_k=q_k).
\end{eqnarray}
Using Bayes' theorem, $ P(q_{k+l}\vert q_k)P(q_k)=P(q_{k+l},q_k)$,
the mean conditional entropy becomes
\begin{eqnarray}
H(Q_{k+l}\vert Q_k) = H(Q_k,Q_{k+l})-H(Q_k),
 				   \label{condentr}
\end{eqnarray}
where the joint Shannon entropy is given by
\begin{eqnarray}
 H(Q_k,Q_{k+l}) = -\sum_{q_k,q_{k+l}}P(q_{k+l},q_k)
\log_2P(q_{k+l},q_k).
\label{shnjoint}
\end{eqnarray}
These Shannon entropies always follow the inequality \cite{bc}
\begin{equation}
H(Q_{k+l} \vert Q_k) \leq H(Q_{k+l}) \leq H(Q_k,Q_{k+l}).
\label{bceqn}
\end{equation}
  The left side of the equation implies
 that removing a constraint never decreases the entropy, and the right side
 implies information stored in two variables is always greater than or equal to 
 that in one \cite{usha}. 
Suppose that three measurements $Q_k$, $Q_{k+l}$, and $Q_{k+m}$, are performed at time instants
$t_k < t_{k+l} < t_{k+m}$. Then,
from equations (\ref{condentr}) and (\ref{bceqn}), the following inequality can 
 be obtained:
\begin{equation}
H(Q_{k+m} \vert Q_k) \leq H(Q_{k+m} \vert Q_{k+l})+H(Q_{k+l} \vert Q_k).
 \end{equation}
For $ n $ measurements $Q_{1},Q_{2}, \dots ,Q_{n}$, at  time instants 
$ t_1 < t_{2} < \dots < t_n $, the above inequality can be 
 generalized to \cite{usha}
\begin{equation}
\sum_{k=2}^nH(Q_{k} \vert Q_{k-1})-H(Q_n \vert Q_1) \geq 0.
\label{elgi}
 \end{equation}
This inequality must be followed by all macro-realistic objects, since 
its satisfaction means the existence of legitimate JP distribution,
which can yield all MP \cite{kurz}.

Usha Devi \textit{et al.} \cite{usha} have shown theoretically that the above 
inequality is violated by a quantum spin-$s$ system, prepared in a 
completely mixed initial state, $ \rho_{in} = \mathbbm{1}/(2s+1)$.
Consider the $ z $-component of the spin evolving under the Hamiltonian ${\cal H}=-\omega S_x $ as 
our dynamical observable, \textit{i.e.} $ Q_t = U_tS_zU_t^\dag $, where
$ U_t=e^{-i{\cal H}t} $, and $S_x$ and $S_z$ are the components of spin-angular momentum.
Let $n$-measurements occur at regular time instants $\Delta t, ~2 \Delta t, \cdots, n\Delta t$.
Ideally in this case, the conditional entropies $ H(Q_k \vert Q_{k-1}) $ between successive measurements 
are all equal, and can be denoted as $ H [ \theta /(n-1) ] $, where $ \theta/(n-1) = \omega \Delta t $
is the rotation caused by the Hamiltonian in the interval $\Delta t$.
Similarly we can denote $ H(Q_n \vert Q_1) $ as $ H [\theta] $.
The lhs of inequality (\ref{elgi}) scaled in units of $ \log_2(2s+1) $ is termed as
the information deficit ${\cal D}$.  For $n$-equidistant measurements, it can be written as
\cite{usha}
\begin{equation}
{\cal D}_n(\theta)=\frac{(n-1)H[\theta / (n-1)] - H[ \theta ]}{\log_2(2s+1)} \geq 0.
\end{equation}

\textit{Measurement of Probabilities.---}
Consider a spin-1/2 particle as the system qubit.
Using the eigenvectors $\{ \ket{0}, \ket{1}\}$ of $S_z$,  as the computational basis,
the projection operators at time $t=0$ are
$\{\Pi_\alpha = \ket{\alpha} \bra{\alpha}\}_{\alpha = 0,1}$.
For the dynamical observable, the measurement basis is rotating under the 
unitary $ U_t=e^{i\omega S_x t} $, such that
$ \Pi_\alpha^t= U_t\Pi_\alpha U_t^\dag$. 
However, it is convenient to perform the actual measurements in 
the time-independent computational basis.
Since for an instantaneous state $\rho(t)$, 
$\Pi_\alpha^t \rho(t) \Pi_\alpha^t = U_t \Pi_\alpha \left( U_t^\dagger \rho(t) U_t \right) \Pi_\alpha U_t^\dagger$,
measuring in $ \{\Pi_\alpha^t\} $ basis is equivalent to back-evolving the state by 
$U_t^\dagger$,
measuring in computational basis, and lastly forward evolving by
$U_t$. This latter evolution can be omitted if one is interested
only in the probabilities and not in the post measurement state of the
system. For example, in case of multiple-time measurements, the forward
evolution can be omitted after the final measurement. 
 
\begin{figure}\centering
\begin{minipage}[t]{.1\textwidth}
\vspace{.1cm}\hspace*{-2cm}(a)
\[ \Qcircuit @C=.3em @R=.3em { \lstick{\rho_S} & \gate{U_i^\dag} & \meter \\ } \]
\end{minipage}
\vrule\hspace{1.2cm}
\begin{minipage}[t]{0.2\textwidth}
\hspace*{-4.3cm}(b)\hspace*{1.2cm}\vspace*{-.6cm}
\[ 
\Qcircuit @C=.3em @R=.3em {
\lstick{\rho_S}  & \qw& \gate{U_i^\dag}  & \multigate{1}{C} & \gate{U_i} & \qw& \qw& \gate{U_j^\dagger}& \qw  & \meter  \\
\lstick{\ket{0}\bra{0}} & \qw& \qw 			                  & \ghost{C} 	&\qw& \qw     & \qw& \qw & \qw & \meter  
\gategroup{1}{3}{2}{5}{.8em}{--}
}
\]~ \\
\vspace*{-0.7cm}
\begin{minipage}{0.1\textwidth} \[ \Qcircuit @C=.8em @R=.8em { &\ctrlo{1}&\qw \\ &\targ&\qw } \] \end{minipage}
\hspace*{-1.2cm}
\begin{minipage}{0.1\textwidth} ~\\~\\ or \end{minipage}
\hspace*{-1.6cm}
\begin{minipage}{0.1\textwidth} \[ \Qcircuit @C=.8em @R=.8em { &\ctrl{1}&\qw \\ &\targ&\qw } \] \end{minipage}
\hspace*{-1.1cm}
\begin{minipage}{0.1\textwidth} ~\\~\\ = \end{minipage}
\hspace*{-1.9cm}
\begin{minipage}{0.1\textwidth} \[ \Qcircuit @C=.6em @R=.6em { &\multigate{1}{C} &\qw \\ &\ghost{C} &\qw} \] \end{minipage}
\end{minipage}
\vspace{.1cm}\hrule\vspace{.2cm}
\begin{minipage}{.25\textwidth}
\hspace*{-5.4cm}(c)\hspace*{1.2cm}\vspace*{-.7cm}
\[
\Qcircuit @C=.45em @R=.45em {
\lstick{\rho_S} &\gate{U_i^\dag}&\qw &\ctrl{1}&\gate{U_i}&\qw&\gate{U_j^\dag}&\qw&\ctrl{2}&\gate{U_j} &\qw& \gate{U_k^\dagger} & \meter\\
\lstick{\ket{0}\bra{0}}&\qw            &\qw               &\targ    &\qw       &\qw&\qw            &\qw&\qw&\qw &\qw&\qw&\meter\\
\lstick{\ket{0}\bra{0}}&\qw            &\qw               &\qw      &\qw       &\qw&\qw            &\qw&\targ   &\qw &\qw&\qw&\meter
\gategroup{1}{2}{2}{5}{.7em}{--}
\gategroup{1}{7}{3}{10}{.7em}{--}
} 
\]
\end{minipage}
\caption{Circuits for measuring SEP (a), two-time JP (b),
and three-time JP (c).
The grouped gates represent measurement in
$\{\Pi_0^t,\Pi_1^t\}$ basis.
In (b) the operation $C$ can be either CNOT or anti-CNOT gate as
described in the text.
}
\label{circuit}
\end{figure}
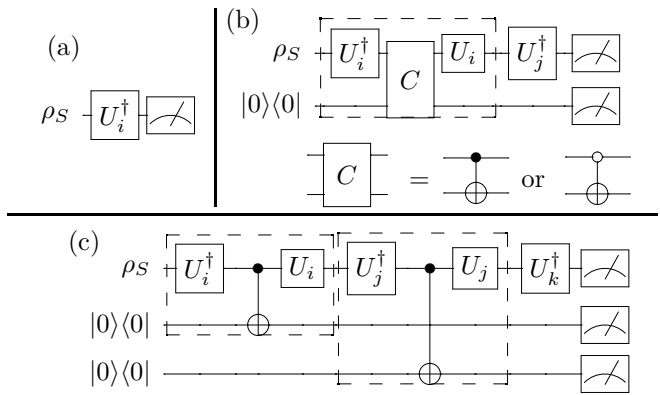

The method for extracting SEP and JP
involves the quantum circuits shown in Fig. \ref{circuit}.
To measure SEP $P(q_i)$ of system qubit in a
general state $\rho_S$, it is evolved by
$U_i^\dagger = e^{i {\cal H} t_i}$, and the probabilities $P(q_i)$ are 
obtained using diagonal tomography (Fig. \ref{circuit}(a)).
Here a further forward evolution by $U_i$ is not necessary as described earlier.

To measure JP $P(q_i,q_j)$, we utilize an ancilla qubit 
initialized in the state $\ket{0}\bra{0}$
(Fig. \ref{circuit}(b)).
After back evolution to computational basis, the CNOT gate encodes the probabilities
of the system-qubit $P(q_i)$ on to the ancilla-qubit \cite{sup}.  After a further evolution by 
$U_i U_j^\dagger = e^{-i\omega S_x(t_j-t_i)}$, a diagonal tomography of the two
qubit system yields $P(q_i,q_j)$  \cite{sup}.

A similar scheme, shown in Fig. \ref{circuit}(c), 
is employed for extracting three-time JP. These circuits can be generalized for 
higher order JP or for spin-$s>1/2$ systems, 
using appropriate ancilla register.

In the earlier LGI experiments, NIM have been performed
by either 
(i) a weak measurement which causes minimum disturbance to the quantum state 
\cite{Palacios,*photons,*semiweak} or 
(ii) initializing the system qubit in a maximally mixed state so that
the system density matrix remains unchanged by the measurements \cite{maheshnmr,*souza}.
Recently however, it was noted by Knee \textit{et al.} that a sceptical macrorealist
is not convinced by either of the above methods \cite{knee}.  Instead, they had proposed a 
`ideal negative result measurement' (INRM) procedure that is more convincingly noninvasive \cite{noninvasive}.  
The idea is as follows.  
The CNOT gate is able to flip the ancilla qubit
only if the system qubit is in state $\vert 1 \rangle$, and does nothing
if the system qubit is in state $\vert 0 \rangle$.
Therefore after the CNOT gate, if
we measure the probability of unflipped ancilla, this corresponds to an `interaction-free'
or NIM of $P(q=0)$. 
Similarly, we can implement an anti-CNOT gate, which
flips the ancilla only if the system qubit is in state $\vert 0 \rangle$, and does nothing
otherwise, such that the probability of unflipped qubit now gives $P(q=1)$.  
Note that in both the cases, the probabilities
wherein the system interacted with the ancilla, resulting in its flip, are discarded.
The final measurement need not be NIM
since we are not concerned about any further evolution. 

In our experiments we combine the two methods, i.e., (i) first we prepare the system 
in a maximally mixed state i.e., $\rho_S = \mathbbm{1}/2$, and 
(ii) we perform INRM.
In this case, $P(0_i) = P(1_i) = 1/2$,
and JP are 
\begin{eqnarray}
&P(0_i,0_j) = |\cos(\theta_{ij}/2)|^2/2 = P(1_i,1_j),& ~~\mathrm{and}, \nonumber \\
&P(0_i,1_j) = |\sin(\theta_{ij}/2)|^2/2 = P(1_i,0_j),&
\label{pqiqj}
\end{eqnarray}
where $ \theta_{ij}=\omega(t_j-t_i)$ 
\cite{usha}.

The only single event entropy needed for the ELGI test is $H(Q_1)$,
since $H(Q_t)$ is constant for the maximally mixed system state.
Further, since $H(Q_1,Q_2) = H(Q_2,Q_3)$ in the case of uniform
time intervals, only two joint entropies $H(Q_1,Q_2)$ and $H(Q_1,Q_3)$
are needed to be measured for evaluating ${\cal D}_3$.
In the following we describe the experimental implementation of these
circuits for the three-measurement LGI test.

\textit{Experiment.---}
We have used $^{13}$CHCl$_3$ (dissolved in CDCl$_3$)
as the two qubit system and treat its $^{13}$C and 
$^1$H nuclear spins as the system and the ancilla qubits respectively.
The resonance offset of $^{13}$C was set to 100 Hz
(provides a dynamic `observable') and
that of $^1$H to 0 Hz (on resonant).  The two spins
have an indirect spin-spin coupling constant $J=209.2$ Hz \cite{sup}.
All the experiments were carried out at an ambient temperature
of 300 K on a 500 MHz Bruker UltraShield NMR spectrometer.

\begin{figure}
\hspace*{-0.7cm}
\includegraphics[width=5.5cm,angle=270]{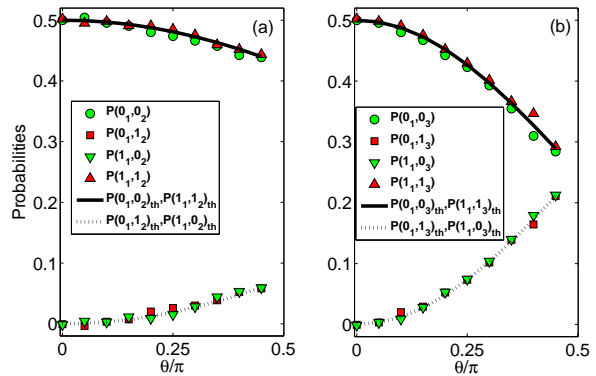}
\caption{The lines indicate theoretical 
JP  and the symbols indicate the mean experimental probabilities
obtained by INRM procedure.
}
\label{prob}
\end{figure}

The evolution propagator $Uj^\dagger U_i = e^{-i S_x \omega (t_j-t_i)}$ 
is realized by the cascade $\mathbb{H}U_d \mathbb{H}$,
where $\mathbb{H}$ is the Hadamard gate, and the delay propagator 
$U_d = e^{-i S_z \omega (t_j-t_i)}$ corresponds to the $\hat{z}$-precession
of the system qubit at $\omega = 2 \pi100$ rad/s resonance
off-set.
The $J$-evolution during this delay is refocused by a $\pi$ pulse
on the ancilla qubit.
The CNOT, $\mathbb{H}$, as well as the $\pi$ pulses are realized by
numerically optimized amplitude and phase modulated RF
pulses, and are robust against the RF inhomogeneity with a average
Hilbert-Schmidt fidelity better than 0.998 \cite{fortunato,maheshsmp,khaneja}.
The final measurement of probabilities are carried out by diagonal
tomography. It involved dephasing all the coherences using a strong 
pulsed field gradient followed by a $\pi/30$ detection pulse. 
The intensities of the resulting spectral lines yielded a traceless
diagonal density matrix, which was normalized and added with identity
matrix to extract the probabilities.
As described in Fig. \ref{circuit}(b),
two sets of experiments were performed, one with CNOT and 
the other with anti-CNOT.
We extracted $P(0,q)$ ($q=\{0,1\}$)
from the CNOT set, and $P(1,q)$ from the anti-CNOT set.
The probabilities thus obtained by INRM 
procedure are plotted in Fig. \ref{prob}.
These sets of experiments also allow us to compare the
results from (i) only CNOT, (ii) only anti-CNOT, and
(iii) INRM procedures.
The joint entropies were calculated in each case using the
experimental probabilities and 
the information deficit (in bits) was calculated using the expression
${\cal D}_3 = 2H(Q_2 \vert Q_1) - H(Q_3 \vert Q_1)$.
The theoretical and experimental values of ${\cal D}_3$ for various rotation
angles $\theta$ are shown in Fig. \ref{results}.  We find a general
agreement between the mean experimental ${\cal D}_3$ values with that of the quantum 
theory.  The error bars indicate the standard deviations obtained by
a series of independent measurements.
According to quantum theory, a maximum violation of ${\cal D}_3 = -0.134$ should occur 
at $\theta = \pi/4$.  The experimental values of ${\cal D}_3(\pi/4)$ are
$-0.141 \pm 0.005$, $-0.136 \pm 0.002$, and $-0.114 \pm 0.027$ for
the CNOT, anti-CNOT, and INRM cases respectively.  Thus in all the
cases, we found a clear violation of ELGI.

\begin{figure}
\centering
\hspace*{-0.7cm}
\includegraphics[width=7.5cm,angle=-90]{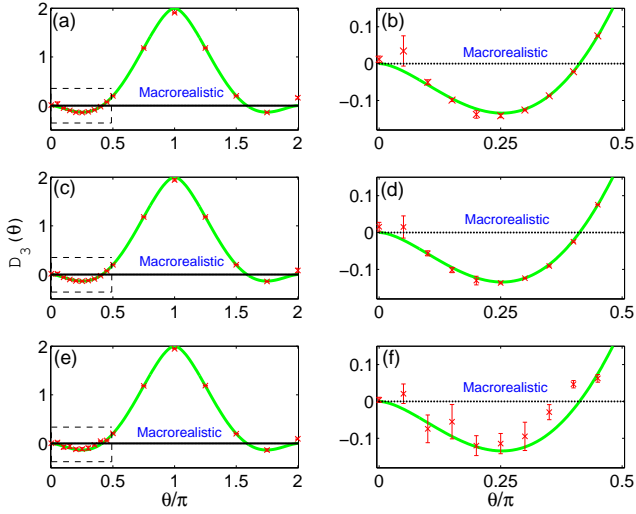}
\caption{
Information deficit ${\cal D}_3$ versus $\theta$ obtained
with CNOT (a,b), anti-CNOT (c,d), and INRM procedure (e,f).
The boxed areas in the left plots (a,c,e) are magnified in the right plots
(b,d,f) respectively.
The mean experimental ${\cal D}_3$ (in bits) values are shown as symbols. 
The curves indicate theoretical ${\cal D}_3$ (in bits).
The horizontal line ${\cal D}_3 = 0$ indicates the lower bound of the 
macrorealism territory.
}
\label{results}
\end{figure}

\textit{Three-time JP.---}
It has been argued that the violation of ELGI in
quantum systems is essentially because certain 
JP can not reproduce all MP \cite{usha}.  For example, in the ${\cal D}_3$ experiment
described earlier, the two-time JP $P(q_1,q_2)$,
$P(q_2,q_3)$, and $P(q_1,q_3)$ can be obtained from eqns. (\ref{pqiqj}).
From the three-time JP, it is possible to generate MP:
\begin{eqnarray}
&P'(q_1,q_2) = \sum_{q_3} P(q_1,q_2,q_3),& \nonumber \\
&P'(q_2,q_3) = \sum_{q_1} P(q_1,q_2,q_3),&~ \mathrm{and} \nonumber \\
&P'(q_1,q_3) = \sum_{q_2} P(q_1,q_2,q_3).&
\end{eqnarray}
Now $P(q_1,q_2,q_3)$ can reproduce $P(q_1,q_2)$ and $P(q_2,q_3)$, i.e., 
$P'(q_1,q_2) = P(q_1,q_2)$ and
$P'(q_2,q_3) = P(q_2,q_3)$.  However, 
$P(q_1,q_2,q_3)$ can not reproduce 
$P(q_1,q_3)$,
i.e., $P'(q_1,q_3) \ne P(q_1,q_3)$,
in general \cite{usha}.  

\begin{figure}[b]
\centering
\hspace*{-0.2cm}
\includegraphics[width=8cm]{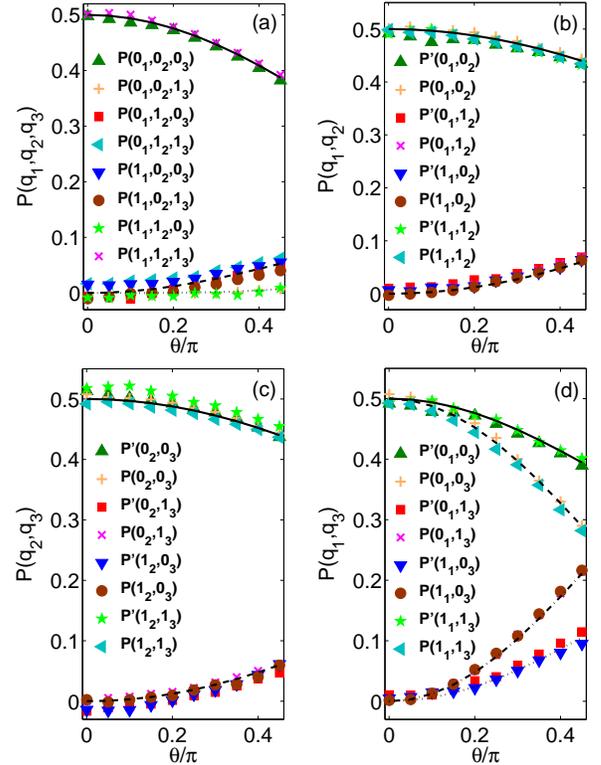}
\caption{
The three-time JP $P(q_1,q_2,q_3)$ (a), and
two-time JP $P(q_i,q_j)$ and
MP $P'(q_i,q_j)$ generated from $P(q_1,q_2,q_3)$ (b-d).
The lines correspond to theoretical values and the symbols
are mean experimental values.
}
\label{p3}
\end{figure}
The above concept can be investigated experimentally
by measuring the three-time JP, as described in Fig. \ref{circuit}(c).
Since this experiment requires measurements at three time instants, we need
two ancilla qubits along with the system qubit. 
We use the three $^{19}$F  nuclear spins (spin-1/2) of 
trifluoroiodoethylene dissolved in acetone-D6
as the three-qubit system \cite{sup}.  
The chemical shifts differences are $\nu_2-\nu_1 = 11860.8$ Hz,
$\nu_2-\nu_3 = 17379.1$ Hz, and the scalar coupling constants 
are: $J_{12} = 69.9$ Hz, $J_{13} = -128.3$ Hz, and $J_{23} = 47.4$ Hz.
Here the first spin (F$_1$) is used as the system qubit 
and the others ($F_2$ and $F_3$) are chosen as the ancilla qubits.
Initialization involved preparing the state,
$\frac{1-\epsilon}{8}\mathbbm{1} + \epsilon \left\{ 
\frac{1}{2}\mathbbm{1}_S \otimes \ket{00}\bra{00}_A \right\}$ where
$\epsilon \sim 10^{-5}$ is the purity factor  \cite{sup,cory}.
The experimental three-time JP $P(q_1,q_2,q_3)$ obtained using the
circuit Fig. \ref{circuit}(c) are shown in Fig. \ref{p3}(a).
Two-time JP $P(q_i,q_j)$ were also obtained using a similar circuit
(Fig. \ref{circuit}(c) without $U_j U_k^\dagger$).
Here JP are completely stored in the ancilla
qubits and were obtained by tracing out the system qubit.
The results $P(q_1,q_2)$, $P(q_2,q_3)$, and $P(q_1,q_3)$ are
shown in Fig. \ref{p3} (b-d).  In each plots, we have overlayed
MP $P'(q_1,q_2)$, $P'(q_2,q_3)$, and $P'(q_1,q_3)$
generated from the experimental three-time JP.  As
expected, MP $P'(q_1,q_2)$ and $P'(q_2,q_3)$ match well 
with the JP $P(q_1,q_2)$ and $P(q_2,q_3)$ respectively, 
while the MP $P'(q_1,q_3)$ does not match with 
JP $P(q_1,q_3)$.  These results confirm that
the three-time JP $P(q_1,q_2,q_3)$ is not 
legitimate in the quantum case.  It is interesting to note
that even for those values of $\theta$ for which ${\cal D}_3$ is positive, 
the three-time JP is illegitimate.  Therefore,
while the violation of ELGI indicates the quantumness
of the system, its satisfaction does not rule out the quantumness.

\textit{Conclusions.---}
We described an experimental study of the entropic Leggett-Garg
inequality in nuclear spins using NMR techniques.
We employed the recently described `ideal negative result measurement' 
procedure to noninvasively extract joint probabilities.
Our results indicate the macrorealistic
bound being violated by over four standard deviations, confirming the
non-macrorealistic nature of the spin-1/2 particles.  
Quantum systems do not have legitimate joint probability 
distribution, which results in the violation of bounds set-up for
macrorealistic systems. 
We have experimentally measured the three-time 
joint probabilities and confirmed that the two-time joint 
probabilities are not reproduced as marginals.

One distinct feature of the entropic LGI is that, the dichotomic
nature of observables assumed in the original formulation of LGI
can be relaxed, thus allowing one to study the quantum behavior 
of higher dimensional systems such as spin $>1/2$ systems.
This could be an interesting topic for future experimental investigations.

The authors are grateful to Prof. Usha Devi, Prof. A. K. Rajagopal, 
Dr. G. C. Knee, Prof. Anil Kumar, Dr. V. Athalye, and S. S. Roy for discussions. 
This work was partly supported by the DST project SR/S2/LOP-0017/2009.

\bibliographystyle{apsrev4-1}
\bibliography{bibelgi}

\newpage

\begin{titlepage}
\section*{Supplementary information for 
``Violation of Entropic Leggett-Garg Inequality in Nuclear Spins"}
\title{Violation of Entropic Leggett-Garg Inequality in Nuclear Spins}
\author{Hemant Katiyar$^1$, Abhishek Shukla$^1$, Rama Koteswara Rao$^2$, and T. S. Mahesh$^1$}

\noindent\textbf{Encoding the probability using a CNOT gate:}
Consider a system qubit initially prepared in a general state $\rho_S$
and an ancilla qubit prepared in the state $\ket{0}\bra{0}$.
The state of the system qubit 
can be written as $P(0_i) \ket{0}\bra{0} + P(1_i) \ket{1}\bra{1} + a \ket{1}\bra{0} + a^\dagger \ket{0}\bra{1}$, where $a$ is a complex probability amplitude.
The CNOT gate encodes the probability of the outcomes in the diagonal elements 
of ancilla qubit since,
\begin{eqnarray}
 ( P(0_i) \ket{0}\bra{0}_S + P(1_i) \ket{1}\bra{1}_S + a \ket{1}\bra{0}_S + a^\dagger \ket{0}\bra{1}_S ) \otimes \ket{0}\bra{0}_A 
\stackrel{\mathrm{CNOT}}{\longrightarrow}
& \ket{0}\bra{0}_S \otimes P(0_i) \ket{0}\bra{0}_A 
+ \ket{1}\bra{1}_S \otimes P(1_i) \ket{1}\bra{1}_A &  \nonumber \\ 
&+ \ket{1}\bra{0}_S \otimes a \ket{1}\bra{0}_A
+ \ket{0}\bra{1}_S \otimes a^\dagger \ket{0}\bra{1}_A. & \nonumber
\end{eqnarray}
The probabilities $P(0_i)$ and $P(1_i)$ can now be retrieved 
by tracing over the system qubit and reading the diagonal elements of
the ancilla state.\\

\noindent\textbf{The qubit systems:}
Fig. \ref{tfie} shows the 
molecular structures and the Hamiltonian parameters of
$^{13}$CHCl$_3$ (Fig. \ref{tfie}(a,b)) and trifluoroiodoethylene (Fig. \ref{tfie}(c,d)).
In the case of $^{13}$CHCl$_3$,
spin-lattice (T1) and spin-spin (T2) relaxation
time constants for the $^{1}$H spin are, respectively, 4.1 s and
4.0 s. The corresponding time constants for $^{13}$C are 5.5 s
and 0.8 s.
In the case of trifluoroiodoethylene,
the effective $^{19}$F transverse relaxation time constants ($T_2^*$) were about 0.8 s and their
longitudinal relaxation time constants were all longer than 6.3 s.\\

\begin{figure}[H]
\centering
\vspace*{-0.5cm}
\includegraphics[width=6.8cm,height=10cm,angle=90]{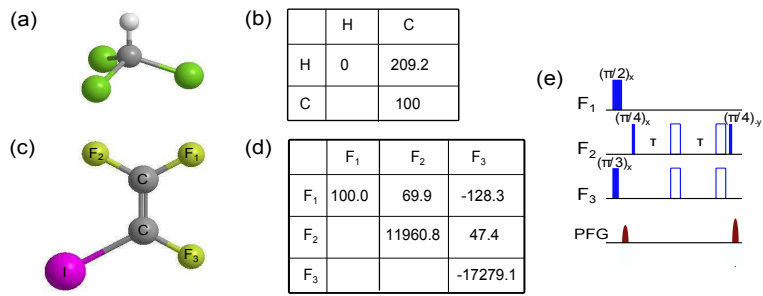}
\vspace*{-3cm}
\caption{The molecular structures of chloroform (a) and trifluoroiodoethylene (c) 
and the corresponding tables (b and d) of relative resonance frequencies 
(diagonal elements) and the J-coupling constants.  The pulse sequence for 
initializing trifluoroiodoethylene is shown in (e).
In (e) the open pulses are $\pi$ pulses and the delay $\tau = 1/(4J_{23})$.}
\label{tfie}
\end{figure}

\noindent\textbf{Initialization in two-qubit ($^{13}$CHCl$_3$) system:}
The initialization involved preparing the maximally mixed state
$\rho_S = \mathbbm{1}/2$ on the system qubit ($^{13}$C).  This
is achieved by a $\pi/2$ pulse on $^{13}$C followed by a 
strong Pulsed Field Gradient (PFG).\\

\noindent\textbf{Initialization in three-qubit (trifluoroiodoethylene) system:}
The equilibrium deviation density matrix
evolves under the PPS sequence (Fig. \ref{tfie}(e)) as follows
\begin{eqnarray}
S_{1z} + S_{2z} + S_{3z}
\stackrel{(\pi/2)_{1x} (\pi/3)_{3x}, \mathrm{PFG}}{\longrightarrow}  S_z^2 + \frac{1}{2}S_{3z} 
\stackrel{(\pi/4)_{2x}}{\longrightarrow} 
\frac{1}{\sqrt{2}}S_{2z} - \frac{1}{\sqrt{2}}S_{2y} + \frac{1}{2}S_{3z} 
\stackrel{1/(2J_{23})}{\longrightarrow} 
&\frac{1}{\sqrt{2}}S_{2z} + \sqrt{2}S_{2x} S_{3z} + \frac{1}{2}S_{3z}& \nonumber \\
&\downarrow& \hspace*{-1.5cm}(\pi/4)_{-2y},\mathrm{PFG} \nonumber \\
&\frac{1}{2} (S_{2z} + 2S_{2z} S_{3z} + S_{3z} ). \nonumber&
\end{eqnarray}

The above deviation density matrix is equivalent to the traceless part of
$\frac{1-\epsilon}{8}\mathbbm{1} + \epsilon \left\{ 
\frac{1}{2}\mathbbm{1}_S \otimes \ket{00}\bra{00}_A \right\}$ where
$\epsilon \sim 10^{-5}$ is the purity factor  \cite{cory}.\\

\noindent\textbf{Diagonal tomography:}\\
The diagonal tomography was
carried out using a strong PFG to dephase out
all the residual off-diagonal elements and using a $6^\circ$ 
non-selective linear readout pulse.  The resulting intensities of
the spectral lines constrain the diagonal elements ($d_i$) of the 
traceless deviation density matrix.  
The experimental deviation density matrix is normalized w.r.t. the
theoretical traceless density matrix such that they both have the same root 
mean square value $\sqrt{\sum_i d_i^2}$, and trace is introduced by adding
identity matrix to the normalized deviation matrix.
The resulting diagonal density matrix yields the probabilities.
Estimation of random errors were carried out by several repetitions at each 
$\theta$ value.\\

\end{titlepage}

\end{document}